\documentstyle[epsf]{l-aa}
\begin{document}

\newcommand{\as}[2]{$#1''\,\hspace{-1.7mm}.\hspace{.1mm}#2$}
\newcommand{\am}[2]{$#1'\,\hspace{-1.7mm}.\hspace{.0mm}#2$}
\newcommand{\dgr}{\mbox{$^\circ$}}    
\newcommand{\E}[1]{\mbox{${}\,10^{#1}{}$}}
\newcommand{\ea}{{\it et al.}}
\newcommand{\grd}[2]{\mbox{#1\fdg #2}}
\newcommand{\gsim}{\stackrel{>}{_{\sim}}}
\newcommand{\lsim}{\stackrel{<}{_{\sim}}}
\newcommand{\Ha}{\mbox{H$\alpha$}}
\newcommand{\HI}{\mbox{H\,{\sc i}}\,\,}
\newcommand{\HIshort}{\mbox{H\,{\sc i}}\,}
\newcommand{\HIbf}{\mbox{H\hspace{0.155 em}{\footnotesize \bf I}}}
\newcommand{\HIit}{\mbox{H\hspace{0.155 em}{\footnotesize \it I}}}
\newcommand{\HIsl}{\mbox{H\hspace{0.155 em}{\footnotesize \sl I}}}
\newcommand{\HIss}{\mbox{H\,{\sc i}}}
\newcommand{\HII}{\mbox{H\,{\sc ii}}}
\newcommand{\kms}{\mbox{\rm km\,s$^{-1}$}}
\newcommand{\kmsMpc}{\mbox{\rm km\,s$^{-1}$\,Mpc$^{-1}$}}
\newcommand{\LB}{\mbox{$L_{B}$}}
\newcommand{\LBnul}{\mbox{$L_{B}^0$}}
\newcommand{\LBsun}{\mbox{$L_{\odot,B}$}}
\newcommand{\Lsun}{\mbox{$L_{\odot}$}}
\newcommand{\LsunMsun}{\mbox{$L_{\odot}$/${\cal M}_{\odot}$}}
\newcommand{\LFIR}{\mbox{$L_{FIR}$}}
\newcommand{\LFIRLB}{\mbox{$L_{FIR}$/$L_{B}$}}
\newcommand{\LFIRLBnul}{\mbox{$L_{FIR}$/$L_{B}^0$}}
\newcommand{\LFIRLsun}{\mbox{$L_{FIR}$/$L_{\odot,Bol}$}}
\newcommand{\MHI}{\mbox{${\cal M}_{\rm HI}$}}
\newcommand{\MHILB}{\mbox{${\cal M}_{\rm HI}$/$L_{\rm B}$}}
\newcommand{\MHILBnul}{\mbox{${\cal M}_{HI}$/$L_{B}^0$}}
\newcommand{\Msun}{\mbox{${\cal M}_\odot$}}
\newcommand{\MsunLsun}{\mbox{${\cal M}_{\odot}$/$L_{\odot,Bol}$}}
\newcommand{\MsunLBsun}{\mbox{${\cal M}_{\odot}$/$L_{\odot,{\rm B}}$}}
\newcommand{\MT}{\mbox{${\cal M}_{\rm T}$}}
\newcommand{\MTLBnul}{\mbox{${\cal M}_{T}$/$L_{B}^0$}}
\newcommand{\MTLBsun}{\mbox{${\cal M}_{T}$/$L_{\odot,B}$}}
\newcommand{\mi}{\mbox{$\mu$m}}
\newcommand{\nan}{\mbox{Nan\c{c}ay}}
\newcommand{\NH}{\mbox{N$_{HI}$}}
\newcommand{\OIII}{\mbox{[O\,{\sc iii}]}}
\newcommand{\s}{\mbox{$\sigma$}}
\newcommand{\Tb}{\mbox{$T_{b}$}}
\newcommand{\tis}[2]{$#1^{s}\,\hspace{-1.7mm}.\hspace{.1mm}#2$}
\newcommand{\vrot}{\mbox{$v_{rot}$}}
\def\la{\mathrel{\hbox{\rlap{\hbox{\lower4pt\hbox{$\sim$}}}\hbox{$<$}}}}
\def\ga{\mathrel{\hbox{\rlap{\hbox{\lower4pt\hbox{$\sim$}}}\hbox{$>$}}}}

\thesaurus{03(11.04.1;   
              11.07.1;   
              11.09.4;   
              13.19.1)}  

\title{An H\,{\large \bf I} line search for optically identified dwarf
  galaxy candidates in the M81 group}

\author{W.~van Driel\,\inst{1}, R.C.~Kraan-Korteweg\,\inst{2}, 
 B.~Binggeli\,\inst{3} \and  W.K.~Huchtmeier\,\inst{4}} 

\offprints{W. van Driel, e-mail : vandriel@mesioq.obspm.fr
\rm}   

\institute{Nan\c{c}ay Radio Observatory, USN, Observatoire de Paris, 
 5 Place Jules Janssen, F-92195 Meudon Cedex, France
\and
DAEC, Observatoire de Paris, 5 Place Jules Janssen, F-92195 Meudon 
 Cedex, France
\and
Astronomical Institute, University of Basel, Venusstrasse 7,
 CH-4102 Binningen, Switzerland
\and
Max-Planck-Institut f\"ur Radioastronomie, Auf dem H\"ugel 69,
 D-53121 Bonn, Federal Republic of Germany} 

\date{Received  14 october 1996; accepted }
\maketitle

\markboth{{W. van Driel et al.: An H\,{\footnotesize I} line search for
optically identified dwarf candidates in the M81 group}}{}

\begin{abstract}
Sensitive 21\,cm \HI line observations were performed for 23 dwarf
members and possible members of the nearby M81 group of galaxies, including
five objects of a clustering of extremely low-surface brightness objects 
of unknown nature. With the \nan\ decimetric radio telescope the 
radial velocity range of --529 to 1826 \kms\ was searched to an rms
noise of $\sim$ 3--5 mJy. Only three objects were detected. However, 
their high radial velocities (between 600 and 1150 \kms) show them to 
lie behind the M81 group. These three objects, classified as 
dS0: (UGC 4998) and Im (Kar 1N and UGC 5658), have \HI\ masses of 
0.5, 2.0 and 2.5$\cdot10^8$\Msun, for the assumed distance of 4 Mpc, 
and \HI\ mass-to-blue light ratios of 0.05, 0.91 and 0.22 \MsunLBsun, 
respectively.

Considering that half of the observed objects are classified as 
irregular dwarfs, hence expected to be relatively gas-rich, the 
resulting detection rate of about $1/3$ is quite low. However,
the mean redshift and velocity dispersion of the M81 group 
($<V> = 101$ \kms, $\sigma=114$ \kms) suggest that the \HI emission 
of low velocity \HIshort-rich members of the M81 group may still remain 
hidden within the strong Galactic \HI emission (typically 
$-150\la V \la 115$ \kms) or, for the 6 dwarf candidates in the
immediate vicinity of M81, overshadowed by the very extended 
\HI envelope encompassing M81, M82, NGC 3077, and NGC 2976 
($-280\la V \la 355$ \kms).

\end{abstract}

\section{Introduction}
Dwarf galaxies provide important clues to the origin and evolution of 
structure in the Universe. Being low-mass objects, dwarfs are most 
vulnerable to interactions with the environment. They are hence ideal 
test particles to study evolutionary processes in different galaxy 
environments and, furthermore, to map the gravitational potential of 
galactic halos, groups and clusters.

Studies of dwarf galaxies have concentrated on the Local Group, and
on clusters such as Virgo and Fornax. Local Group dwarfs can be studied in 
great detail, but there are only a few of them. The clusters, albeit
rich in dwarfs, are relatively distant.  Here, the M81 group provides
the ideal probe: it has about three times the dwarf content of the
Local Group but is at only about a quarter of the Virgo cluster
distance.

An extensive survey of M81 group dwarfs has been carried out 
by B\"orngen et al. 
(1982), resulting in a list of dwarf members to a limiting absolute
magnitude of $\approx -11$, for the adopted distance of 4 Mpc to the group. 
They furthermore report the clustering of 
a number of unusual, very low-surface brightness about 7\dgr\
southeast of M81. Although classified as dwarf members of the M81
group (B\"orngen et al. 1984), the nature of these objects is
uncertain.  Are these remarkable objects dwarfs at the 
extreme faint end of the luminosity function? And if true, do
they contain any gas and might thus be the lowest \HIshort-mass objects 
yet observed?

The goal of the survey presented here is to determine the
\HI properties of the dwarfs of the M81 group. Dwarf galaxies in
groups and in the field generally are irregulars (Binggeli et
al. 1990) -- hence gas-rich. We therefore searched at \nan\ for 
21\,cm \HI line emission in 23 candidate dwarf galaxy members of 
the M81 group with considerably lower rms noise than previously obtained.

%
%
\begin{table*}
\bigskip
\small
\begin{tabular}{crlcclrcrcc}
\multicolumn{11}{l}{{\bf Table 1:} Members and possible members (P) of
  the M81 group of galaxies} \\
\smallskip \\
\hline
\vspace{-3 mm} \\
 & & Ident. & R.A. & Dec. & Type & Diam & $B_{\rm T}$ & 
$V_{\rm hel}$
 & \HI\ &\nan \\
 & & & (h\,\, m\,\, s) & (\degr\,\, \arcmin\,\, \arcsec) & 
& \,(\,\arcmin\,) & (mag) & (km/s) & \\
\vspace{-3 mm} \\
\hline
\vspace{-2 mm} \\
 & 1.  & Ho\,{\sc ii}      & 08 13 53.5 & 70 52 13 & Im     &  7.9 & 10.2 & 157$\pm$1  &
 det & \\
 & 2.  & Kar 52    & 08 18 43.0 & 71 11 25 & Im     &  1.3 & 14.2 & 114$\pm$6  &
 det & \\
 & 3.  & DDO 53    & 08 29 33.3 & 66 21 08 & Im     &  1.5 & 13.6 &  19$\pm$10  &
 det & \\
 & 4.  & UGC 4483  & 08 32 07.0 & 69 57 16 & Im     &  1.1 & 13.9 & 156$\pm$5  &
 det & \\ 
 & 6.  & UGC 4998  & 09 20 52.9 & 68 35 53 & dS0:   &  1.6 & 13.9 &      &
 no det & * \\
 & 7.  & Ho\,{\sc i}      & 09 36 00.9 & 71 24 55 & Im     &  3.6 & 12.2 & 136$\pm$3  &
 det & \\
P& 8.  & Kar 1N    & 09 41 00.0 & 69 37 00 & Im     &  1.2? & 15.5 &     &
 & * \\
 & 9.  & N 2976    & 09 43 11.5 & 68 08 45 & Sd     &  5.9 & 10.9 &   3$\pm$5  &
 det & \\
 &10.  & Kar 2N    & 09 43 42.0 & 69 30 00 & dE?    &      & 15.9 &      &
 no det & *  \\
 &11.  & Kar 59    & 09 46 40.0 & 72 17 41 & Im     &  0.3 & 17.1 &      &
 no det & * \\
 &12.  & Kar 3N    & 09 49 42.0 & 69 12 00 & Im     &  0.5 & 17.1 & $-$40$\pm$60  &
 no det & * \\
 &13.  & M 81      & 09 51 27.3 & 69 18 08 & Sb     & 26.9 &  7.9 & $-$34$\pm$4  &
 det & \\
 &14.  & M 82      & 09 51 43.6 & 69 55 00 & Amorph & 11.2 &  9.3 & 203$\pm$4  &
 det & \\
 &15.  & A952+69   & 09 53 27.0 & 69 31 18 & Im     &      & 14.3 &      &
 & * \\
 &16.  & Kar 61    & 09 53 01.0 & 68 49 48 & dE,N   &  1.2 & 13.8 &      &
 & *  \\  
 &17.  & Ho {\sc ix}     & 09 53 28.0 & 69 16 53 & Im     &  2.5 & 13.5 &  46$\pm$6  &
 det & \\     
 &18.  & NGC 3077  & 09 59 21.9 & 68 58 33 & Amorph &  5.4 & 10.6 &  14$\pm$4  &
 det & \\
 &19.  & Garland   & 09 59 54.0 & 68 55 30 & Im     &      &      &  50  &
 & * \\
 &20.  & Kar 5N    & 10 00 42.0 & 68 30 00 & dE:    &      & 18.1 &      &
 no det & * \\
 &21.  & DDO 71    & 10 01 18.0 & 66 47 53 & dE,N   &  0.9 & 14.6 &    &
 no det & * \\ 
P&22.  & UGC 5423  & 10 01 25.3 & 70 36 27 & BCD    &  0.9 & 13.8 &  349$\pm$5 &
 det & \\
 &23.  & Kar 64    & 10 03 07.2 & 68 04 20 & dE,N   &  1.8 & 14.9 &      &
 no det & * \\
 &24.  & DDO 78    & 10 22 48.0 & 67 54 40 & dE     &  2.0 & 14.3 &      &
 no det & * \\
P&25.  & UGC 5658  & 10 23 52.6 & 71 29 34 & Im     &  1.0 & 15.0 &      &
 no det & * \\
 &26.  & IC 2574   & 10 24 41.3 & 68 40 18 & Sm    & 13.2  & 11.0 &   47$\pm$3 &
 det & \\
 &27.  & DDO 82    & 10 26 48.0 & 70 52 33 & Sm/BCD: & 3.2 & 11.8 &   40 &
 no det & * \\
 &28.  & Kar 6N    & 10 31 00.0 & 66 16 00 & dE     &     & 16.0 &       &
 no det & * \\
P&29.  & Anon 1    & 10 45 30.0 & 65 02 00 & ?      &      & 17.4 &      &
 &  \\
P&30.  & UGCA 220  & 10 46 04.0 & 64 59 00 & Im:    &  1.7 & 16.9 &      &
 no det & * \\
P&31.  & DDO 87    & 10 46 17.0 & 65 47 40 & Im:    &  2.4 & 14.9 &  338$\pm$5 &
 det & \\
P&32.  & Anon 2    & 10 46 48.0 & 65 00 00 & ?      &      & 16.3 &      &
 &   \\
P&33.  & Kar 7N    & 10 47 06.0 & 65 22 00 & ?      &      & 16.2 &      &
 no det & * \\
P&34.  & Anon 3    & 10 47 12.0 & 65 00 00 & ?      &      & 16.5 &      &
 &   \\
P&35.  & Anon 4    & 10 47 18.0 & 65 00 00 & ?      &      & 16.5 &      &
 \\
 &36.  & Kar 73    & 10 49 30.0 & 69 48 55 & Im     &  0.6 & 14.9 &  115 &
 no det & * \\
P&37.  & Anon 5    & 10 50 30.0 & 65 31 00 & ?      &      & 15.5 &      &
 & * \\
P&38.  & Anon 6    & 10 50 54.0 & 65 17 00 & ?      &      & 16.1 &      &
 & * \\
P&39.  & Kar 8N    & 10 51 06.0 & 65 28 00 & ?      &      & 15.4 &      &
 & * \\
P&40.  & Anon 7    & 10 51 18.0 & 65 33 00 & ?      &      & 15.8 &      &
 & * \\
 &41.  & Kar 74    & 10 59 05.2 & 70 32 01 & dE/Im: &  1.0 & 15.2 &      & 
 no det & * \\
\vspace{-2 mm} \\
\hline
\vspace{-2 mm}\\
\multicolumn{11}{l}{\footnotesize {\bf Note:} Column 1: P denotes a possible 
 group member; Column 10: indicates objects previously observed in \HI} \\
\multicolumn{11}{l}{\footnotesize and status, i.e. detection or
 no detection; Column 11: * marks dwarfs observed in present survey
(see Table 2)} \\
\end{tabular}
\normalsize
\end{table*}

\section{The M81 group}
The M81 group of galaxies is the most nearby rich concentration of dwarf
galaxies beyond the Local Group. 
For our study of possible dwarf members of the M81 group of galaxies, we
used the optically selected catalog of 41 members and possible members
(dwarfs or otherwise) compiled by Binggeli (1993). 
These data represent a compilation of various publications, databases and 
private communications. Binggeli's catalog is strongly based on the
2 m Tautenberg Schmidt survey by B\"orngen \& Karachentseva (1982),
the photometric work by B\"orngen et al. (1982), and the photographic 
atlas of Karachentseva et al. (1985a) obtained with the 6 m SAO telescope.
Note that we rejected object no. 5 (Kar 54 = UGC 5954, at 
$\alpha$ = 09$^{\rm h}$17\fm6, $\delta$ = 75\degr57\arcmin) due to its 
high redshift of 659 \kms, which was unknown in 1993. We also 
rectified the identification of Ho~{\sc ix}, erroneously named 
Ho~{\sc iv} in Binggeli (1993).

The basic optical data of these objects are listed in Table 1. 
Binggeli's numbering was retained throughout this paper. 
The coordinates were taken from NED and are for the epoch 1950.0. 
All velocities are heliocentric and calculated according to the 
conventional optical definition ($v$ = c$\Delta\lambda/\lambda_{0}$).
Heliocentric velocities are from de Vaucouleurs et al. (1990, RC3), except 
for the following objects: no. 1 Holmberg\,{\sc ii} (Strauss et al. 1992), 
no. 12 Kar 3N (Tikhonov \& Karachentsev 1993), no. 19 Garland 
(Karachentseva et al. 1985b), no. 22 UGC 5423 (Schneider et al. 1992) 
and no. 36 Kar 73 (Tikhonov \& Karachentsev 1993); note that the 
heliocentric velocity of 180 km/s listed for no. 27, DDO 82, in
the RC3 is incorrect (see Sect. 4.1).

%
 \begin{figure*}
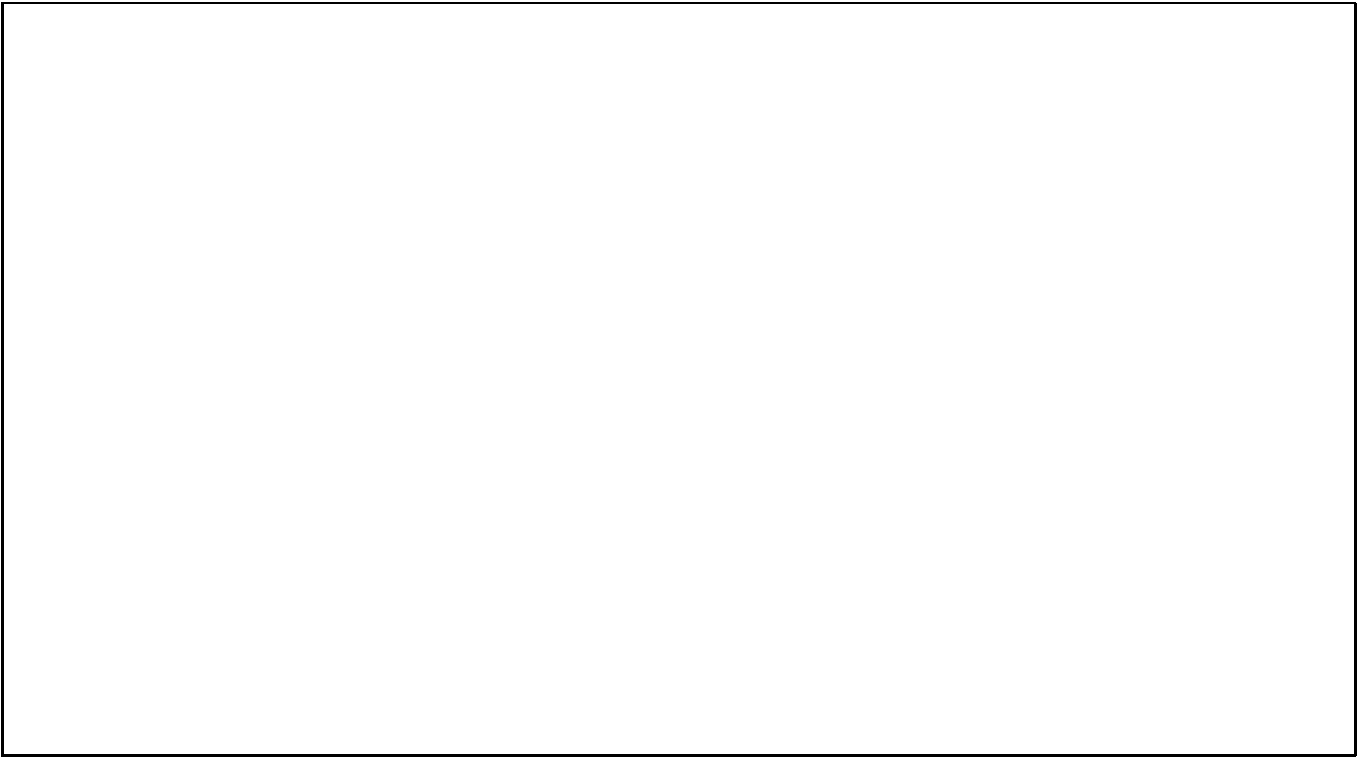
 
\picplace {10cm}
 \caption[]{Distribution on the plane of the sky of all 40 members and
 possible members of the M81 group of galaxies listed in Table 1, within the
 boundaries of the Schmidt survey of B\"orngen \& Karachentseva (1982). 
 The morphological type of each galaxy has been indicated. Uncertain
 group members are put in parentheses, the latter also include the clustering
 of low-surface brightness objects of unknown nature in the SE corner. 
 The names of 
 galaxies previously  detected in \HI have been underlined. The linear
 scale bar of 0.5 Mpc is based on the assumed distance of 4 Mpc.
 Also shown is the HPBW of the \nan\ radio telescope.
 }
 \end{figure*} 

The distribution on the sky of all 40 members and possible members of
the group listed in Table 1 is shown in Figure 1, where the
morphological type of each object has been indicated.
We have adopted a distance of 4 Mpc for the M81 group. This is 
significantly smaller than the value of 5.5 Mpc given in the RSA (Sandage \&
Tammann 1987), but is in good accord with more recent work
(Karachentsev 1996, cf.~his Table 1). 

The M81 group, like others, shows a clear morphological segregation:
most early-type dwarfs are found in a dense core around M81, while the dwarf 
irregulars are spread out over the entire survey area.

The M81 group is more compact than the Local Group, and the core galaxies 
are known to be strongly interacting, as shown by the large, dynamically 
complex \HI cloud embedding M81, M82, NGC 3077, and NGC 2976 (van der 
Hulst 1977, Appleton et al. 1981, recent VLA results in Yun et
al. 1994, and also the dynamical analysis by Karachentsev 1996).
Because of this, and the lack of radial velocities, it is impossible
to determine which dwarf belongs to which specific large galaxy.

The M81 group may well be more dynamically evolved than the Local
Group, having already released its formerly bound dwarfs through 
dynamical friction. On the other hand, the Garland system is 
a knotty dwarf irregular that apparently formed only recently from a 
tidal tail, a system that is likely to have been born free.
A peculiar case is that of the clustering of low-surface brightness 
objects (Kar~7N, Kar~8N, and Anon 1 to 7 in Table 1) 
noted by B\"orngen et al. (1984). 
It is not clear whether these are of an extragalactic nature; they may be 
Galactic cirrus clouds; the M81 group is an area of the sky rife 
with potential for confusion between Galactic and extragalactic objects 
(cirrus, HVCs, tidal tails).

\section{\HI line observations of M81 group dwarfs}

Of the 40 objects listed in Table 1, only 13 have previously been 
detected in the 21\,cm \HI line (see also Table 3). 
Four galaxies (Kar 1N, A952+69, Kar 61, and Garland), as well as 8
of the peculiar 
low-surface brightness objects first noted by B\"orngen
et al. (1984) have never before been observed in the 21\,cm line. These 4
galaxies, as well as the four brightest objects of unknown nature, 
i.e. Kar~8N, and Anon 5, 6 and 7, were selected for our deep \HI\ search.

Four objects (Holmberg\,{\sc i}, Kar 52 (= M81 dwarf A), Kar 73 and Kar 3N)
were mapped in \HI\ with the VLA (Westpfahl \& Prugniel 1994). The first two 
look like rather face-on incomplete rings with evidence for slow rotation,
while the latter two resemble discs with a central hole seen edge-on and
complex or confused velocity fields without clear signs of systematic 
rotation in the dwarf systems. No \HI\ profile parameters are given in this
reference -- note, that neither Kar 73 nor Kar 3N has been reported as
detected in any single-dish study, including ours.

Fifteen of these galaxies have previously been searched for with the 
90m Green Bank (Schneider et al.~1992) and the 100m Effelsberg dish 
(Huchtmeier \& Skillman 1994, 1997), with an rms noise varying from 7 to 
15 mJy, and 4 with an rms noise of 18 mJy only for earlier work by
Fisher \& Tully (1981) at Green Bank and Effelsberg. We reobserved 
these galaxies at \nan\ with considerably higher sensitivity 
(3-5 mJy rms). 

The total sample of 23 dwarf galaxy candidates selected for
observation in \HI at \nan\ are marked in the last column of Table 1.
The searches for 21\,cm \HI line emission were made 
with the \nan\ decimetric radio telescope in the period December 1995 -- 
September 1996. The telescope is a meridian transit-type instrument,
which permits the tracking of such high-declination galaxies for about
3 hours per day. The telescope has the equivalent collecting surface of 
a 94 m diameter round dish, but its HPBW is about 4$'$ $\times$ 22$'$ 
($\Delta\alpha \times \Delta\delta$), due to its elongated geometry. 
Most objects were observed for about 4 hours each, while the 3 low-surface
brightness Anonymous objects were observed for about 8 hours each.
It should be noted that the telescope is somewhat less sensitive at high 
declinations, like those observed for the present survey (65\dgr--72\dgr), 
then at lower declinations, due to its geometry.

The 1024 channel correlator set-up used permitted an \HI line search in both
$H$ and $V$ polarisation in the radial velocity range of --529 to 1826
\kms, with a 190 \kms\ overlap between the filter banks, from 548 to 
747 \kms, at a resolution of about 6 \kms.

After averaging the individual spectra, the data was smoothed to a velocity 
resolution of 12.7 \kms, and third-order baselines were fit. For the 
conversion of antenna temperature to flux density in mJy 
we used the standard calibration relation established by the \nan\ staff
through regular monitoring of strong continuum sources, and the
long-term standard galaxy \HI line monitoring data made available to us  
prior to communication by G. Theureau (see Theureau et al. 1997).

\section{Results}

The resulting \HI spectra of all 23 M81 group candidate dwarf galaxy members
are shown in Fig. 2. 

%
 \begin{figure*}
\hfil \epsfxsize 18cm \epsfbox{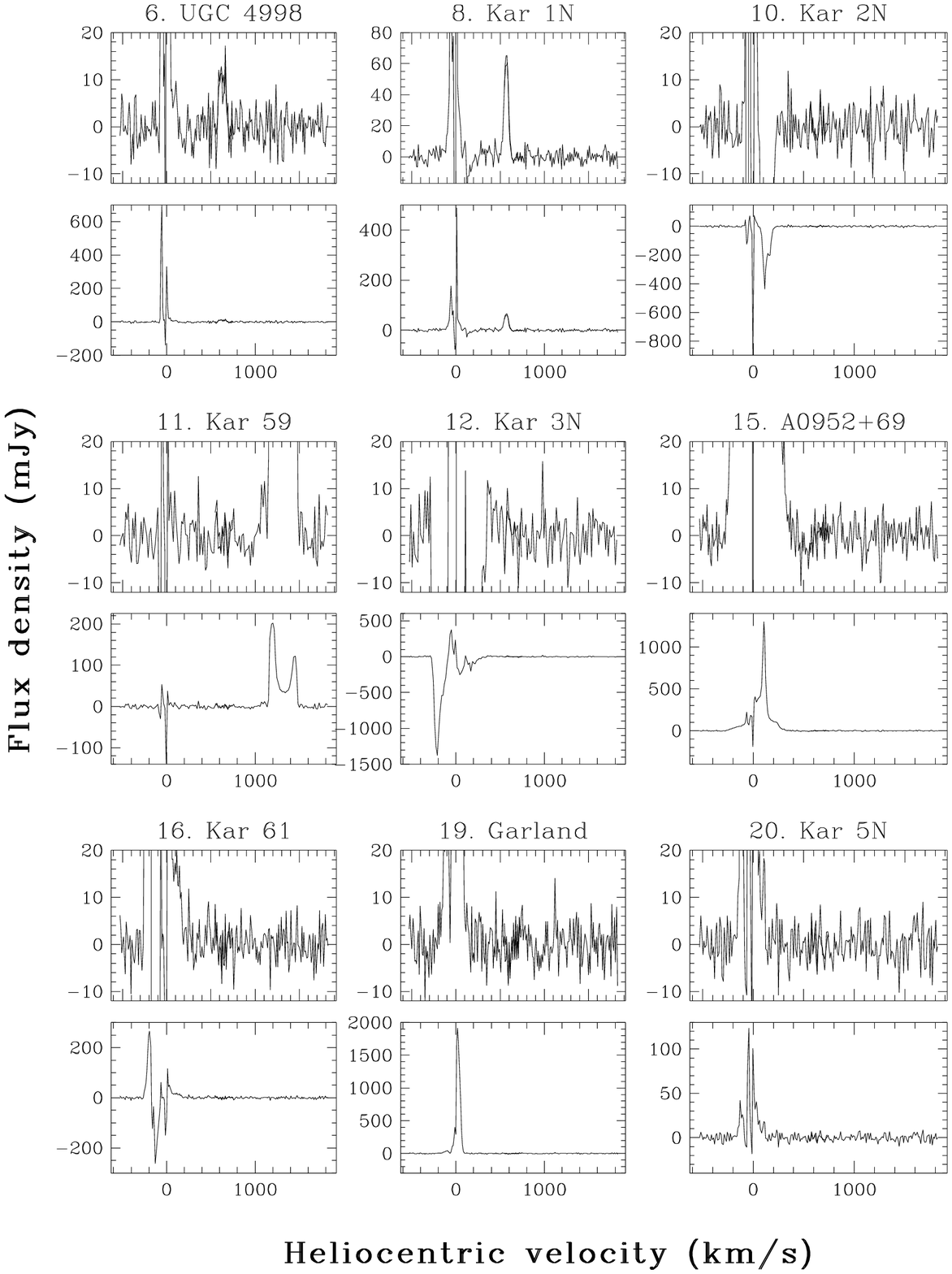}\hfil \clearpage
 \caption[]{Explanations at the end.}
\end{figure*}

\begin{figure*}
\addtocounter{figure}{-1}
\hfil \epsfxsize 18cm \epsfbox{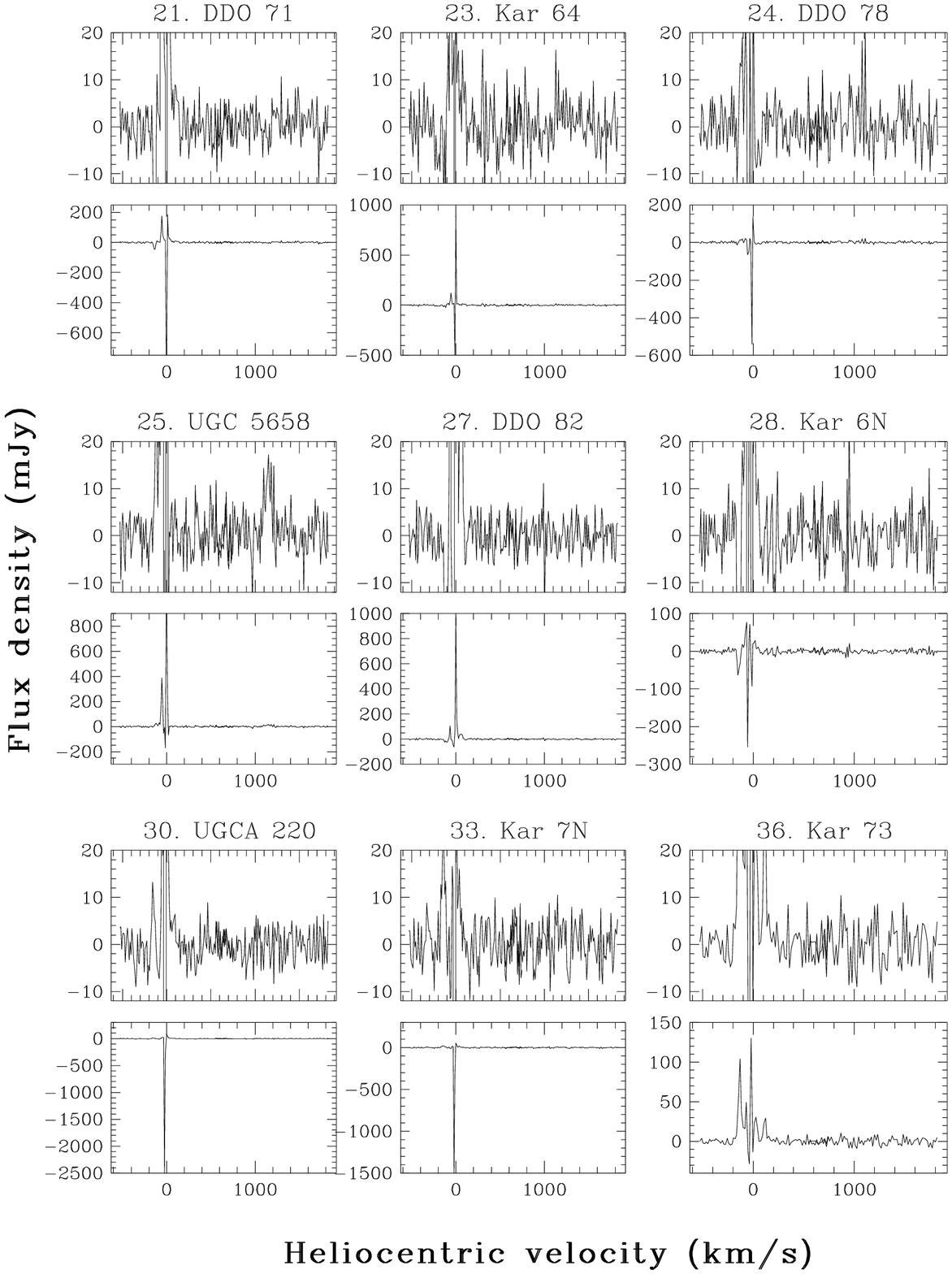}\hfil \clearpage
\caption[]{Explanations at the end.}
\end{figure*}

\begin{figure*}
\addtocounter{figure}{-1}
\hfil \epsfxsize 18cm \epsfbox [20 273 592 779]{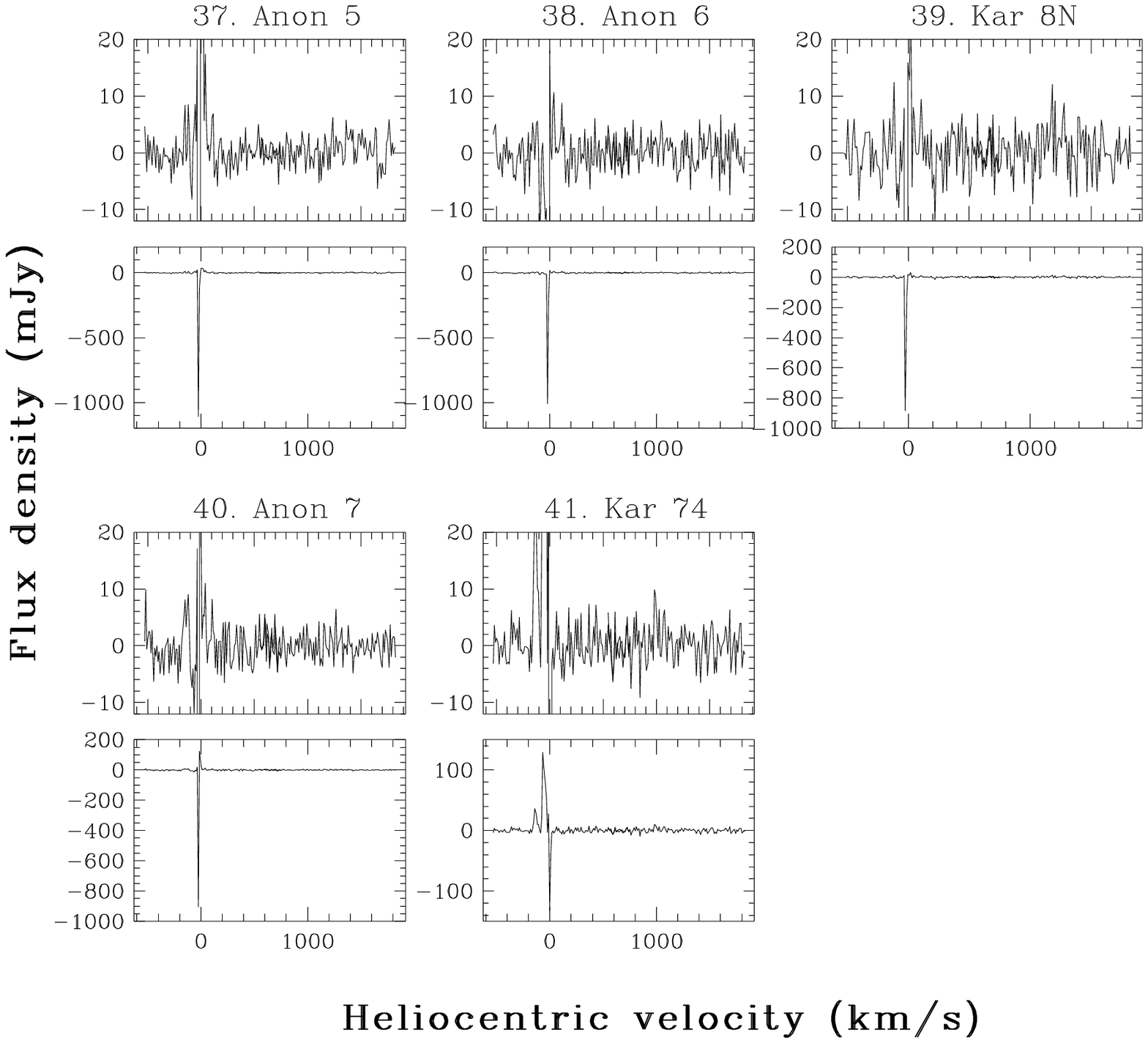}\hfil
  \caption[]{21\,cm \HI line spectra  of all 23 candidate dwarf galaxy 
members of the M81 group observed. Velocity resolution is 12.7 \kms. 
For each object two panels are shown, with the upper displaying a small
range in flux density to reveal faint features, and the lower giving
the full intensity range to demarcate the Galactic \HI features
(typically  at $-150\la V \la 115$ \kms) as well as confusion in the 
\HI complex surrounding M81, in the range --280 \kms\ to +355 \kms, cf. Yun
et al. 1994). Radial velocities are  according to the radio convention.
}
\end{figure*} 

The optical centre positions to which the telescope was pointed, the
rms noise of the spectra at 12.7 \kms\ resolution, as well as
derived \HI line properties are listed in Table 2.
Upper limits to the \HI line flux are 3$\sigma$ values for an assumed 
line width of 50 \kms. The last two columns indicate the velocity range
within which the strong Galactic \HI line signal prevents detection of
external galaxies (typically $-150\la V \la 115$ \kms), or the strong
emission of the interconnected \HI\ distribution around the galaxies 
M81, M82, NGC 3077 and possibly NGC 2976 obliterates the lower flux 
emission of the 6 dwarf candidates in that vicinity, i.e. a
detection of Kar 2N, Kar 3N, A952+69, Kar 61, Garland and Kar 5N 
in the velocity range --280 \kms\ to +355 \kms\ (cf. van der Hulst 1977,
Appleton et al. 1985, and Yun et al. 1994 for detailed \HI maps of this 
complex region).

%
%
\begin{table*}
\bigskip
\footnotesize
\begin{tabular}{rlcclrrlrrrrrr}
\multicolumn{14}{l}{{\bf Table 2:} \nan\ \HI\ observations of 
 the 23 possible dwarf galaxy members of the M81-group} \\
\smallskip \\
\hline
\vspace{-3 mm} \\
 & Ident. & R.A. & Dec. & Type & $B_{\rm T}$ & $V_{\rm opt}$ & rms 
  & $V_{\rm HI}$ & $\int S$d$V$ & $\Delta V_{50}$ & $\Delta V_{20}$ 
  & \multicolumn{2}{c}{Conf. range} \\
 & &  (h\,\, m\,\, s) & (\degr\,\, \arcmin\,\, \arcsec) &  & {\tiny (mag)} 
  & {\tiny (km/s)} & {\tiny (mJy)} & {\tiny (km/s)} 
  & {\tiny (Jy~km/s)} & {\tiny (km/s)} & {\tiny (km/s)} 
  & \multicolumn{2}{c}{\tiny (km/s)} \\
\vspace{-3 mm} \\
\hline
\vspace{-2 mm} \\
6.  & U 4998  & 09 20 53 & 68 35 53 & dS0:   & 13.9 &      & 3.9  &  632 
&    0.84 &  89 & 105 &  --90 & 140 \\  
8.  & Kar 1N  & 09 41 00 & 69 37 00 & Im     & 15.5 &      & 4.4  &  570 
&    3.81 &  56 &  81 & --120 &  75 \\  
10. & Kar 2N  & 09 43 42 & 69 30 00 & dE?    & 15.9 &      & 4.4  &      
& $<$0.65 &     &     & --115 & 250 \\  
11. & Kar 59  & 09 46 40 & 72 17 41 & Im     & 17.1 &      & 5.0  & (1322 
& 34.6  & 311  & 326) & --110 & 100 \\  
12. & Kar 3N  & 09 49 42 & 69 12 00 & Im     & 17.1 &  --40 & 5.0  &      
& $<$0.75 &     &     & --280 & 290 \\  
15. & A952+69 & 09 53 27 & 69 31 18 & Im     & 14.3 &      & 3.9  &      
& $<$0.59 &     &     & --270 & 355 \\  
16. & Kar 61  & 09 53 01 & 68 49 48 & dE,N   & 13.8 &      & 4.2  &      
& $<$0.62 &     &     & --260 & 190 \\  
19. & Garland & 09 59 54 & 68 55 30 & Im     &      &   50 & 4.8  &      
& $<$0.72 &     &     & --170 &  75 \\  
20. & Kar 5N  & 10 00 42 & 68 30 00 & dE:    & 18.1 &      & 4.5  &      
& $<$0.67 &     &     & --195 & 130 \\  
21. & DDO 71  & 10 01 18 & 66 47 53 & dE,N   & 14.6 &      & 4.2  &      
& $<$0.62 &     &     & --160 & 150 \\  
23. & Kar 64  & 10 03 07 & 68 04 20 & dE,N   & 14.9 &      & 5.4 &      
& $<$0.81 &     &     & --115 & 130 \\  
24. & DDO 78  & 10 22 48 & 67 54 40 & dE     & 14.3 &      & 6.1  &      
& $<$0.92 &     &     & --145 &  90 \\  
25. & U 5658  & 10 23 53 & 71 29 34 & Im     & 15.0 &      & 5.0  & 1159 
&    1.50 & 127 & 203: & --155 & 35 \\  
27. & DDO 82  & 10 26 48 & 70 52 33 & Sm     & 11.8 &   40 & 4.2  &      
& $<$0.62 &     &     & --135 & 100 \\  
28. & Kar 6N  & 10 31 00 & 66 16 00 & dE     & 16.0 &      & 5.5 &      
& $<$0.82 &     &     & --170 &  65 \\  
30. & UA 220  & 10 46 04 & 64 59 00 & Im:    & 16.9 &      & 3.9  &      
& $<$0.59 &     &     & --180 & 115 \\  
33. & Kar 7N  & 10 47 06 & 65 22 00 & ?      & 16.2 &      & 5.4  &      
& $<$0.81 &     &     & --175 &  75 \\  
36. & Kar 73  & 10 49 30 & 69 48 55 & Im     & 14.9 &  115 & 5.4  &      
& $<$0.81 &     &     & --165 & 165 \\  
37. & Anon 5  & 10 50 30 & 65 31 00 & ?      & 15.5 &      & 2.9  &      
& $<$0.42 &     &     & --100 &  80 \\  
38. & Anon 6  & 10 50 54 & 65 17 00 & ?      & 16.1 &      & 3.5  &      
& $<$0.51 &     &     &  --55 &  50 \\  
39. & Kar 8N  & 10 51 06 & 65 28 00 & ?      & 15.4 &      & 4.5  &      
& $<$0.67 &     &     &  --50 &  45 \\  
40. & Anon 7  & 10 51 18 & 65 33 00 & ?      & 15.8 &      & 2.8  &      
& $<$0.40 &     &     & --165 &  70 \\  
41. & Kar 74  & 10 59 05 & 70 32 01 & dE/Im: & 15.2 &      & 3.9  &      
& $<$0.59 &     &     & --165 &  60 \\  
\vspace{-2 mm} \\
\hline
\vspace{-2 mm}\\
\multicolumn{14}{l}{\footnotesize {\bf Note:} The strong `Kar 59'
  detection is due to the nearby spiral NGC 2985}\\
\end{tabular}
\normalsize 
\end{table*}

\subsection{Notes to individual galaxies}
In the search for objects that may possibly have confused the 4 
\HI line spectra in which the profile of an external galaxy was detected
(i.e., of UGC 4998, Kar 1N, Kar 59 and UGC 5658) we queried the NED 
database in an area of $6'\times 33'$ 
($\Delta\alpha \times \Delta\delta$), i.e. 1.5 times the HPBW in both
R.A. and Dec.

{\bf No. 6 = U 4998}\, 
A photograph is displayed in the photographic 
atlas of Karachentseva et al. 1985a (forthwith KKB85), obtained with
the 6 m SAO telescope, who suggest a dE7
classification. The structural parameters from photometry based on
scans of plates from the 2 m Tautenburg Schmidt telescope and the 
above KKB85 atlas are given in Karachentseva et
al. (1987), henceforth KKRBF87. 
The \nan\ observations reveal a clear detection at 631 \kms.
The earlier \HI observations by Fisher \& Tully (1981, henceforth
FT81) were not sensitive enough and the signal discovered here is 
outside the velocity range covered in the Green Bank 90-m
\HI\ survey of Schneider al. 1992 (henceforth STMM92).
Within the search area no other galaxies are known, only a 
radio source (87GB 092050.8+68181) at 17\farcm7 distance. 
The detection at \nan\ at 631 \kms\ hence suggests that this dwarf
candidate is not a member of the M81 group but lies
behind the M81 group. 

{\bf No. 8 =  Kar 1N = Mailyan 45}\, The positional agreement 
-- later confirmed by Madore et al.~(1994) -- and corresponding
dimensions suggest this galaxy to be the dwarf Mailyan 45 discovered earlier
(cf. Mailyan Dwarf Galaxy Catalog, Mailyan 1973). 
A photograph and isodensity map is given in KKB85 who suggest that 
this very elongated object might be interacting with the intergalactic
medium, as its light distribution is noticeably lopsided. The \nan\
observations find a clear detection at 569 \kms. No objects which 
might cause confusion were found in NED, 
only a radio source (87GB 094123.3+692131) at 15\farcm7 distance, 
and a faint 60 $\mu$m source (IRAS F09413+6914) at a distance of 
22\farcm2. This irregular galaxy is thus more distant than the M81 group.

{\bf No. 10 = Kar 2N}\, Photograph in KKB85. No structural details can
be seen. Note the strong negative (residual) emission between 0 and 
200 \kms\ in the spectrum given in Fig.~2. This emission is due to 
the higher velocity gas of the western part of M81 entering through
the sidelobes of the \nan\ radio telescope. If this object were to 
have some gas in this velocity range it would not be detectable with 
these observations. 

{\bf No. 11 = Kar 59}\, 
Photograph of this very faint low-surface brightness object
is given in KKB85, its structural parameters in KKRBF87.
Our \HI spectrum shows higher noise in the 170 to 250 \kms\ range, 
due to interference in the $H$ polarization during these observations;
but the other data ($V$ polarization) do not show a detection in this 
velocity range either. The strong, double-horned profile visible in 
the spectrum at 1323 \kms\ is due to NGC 2985, a 11.2 mag Sab spiral 
at 13\farcm5 distance ($\alpha = 09^{\rm h}45^{\rm m}$52\fs6, 
$\delta = 72\degr30\arcmin45\arcsec$), just outside the area searched in NED 
(in which no objects were found). For an \HI line profile of 
NGC 2985, see Oosterloo \& Shostak (1993). KKRBF87 classify this
object as a dwarf spheroidal rather than an irregular which might 
explain the non-detection in \HI. 

{\bf No. 12 = Kar 3N}\, 
The photograph by KKB85 and Tikhonov \& Karachentsev (1993) shows 
resolution into blue stars. The latter determined a photometric 
distance of 2.8 Mpc from the brightest stars. KKB85 consider it to 
form one tidally-disrupted object together with two small fragments 
about 1$'$--2$'$ north of it. Kar 3N might well be \HIshort-rich, but 
this irregular dwarf lies within the extended \HI envelope of 
M81 (cf. Yun et al. 1994) and any emission from the
dwarf within $-270 \la V \la$ 355 \kms will be lost in the strong 
and broad emission from M81 itself (cf. Fig.~2).
VLA \HI\ imaging (Westpfahl \& Puche 1994) shows a structure like
an edge-on disc with a central hole; the velocity field shows systematic 
rotation, but it is not clear if this is due to the local M81 arm or
to internal motions in Kar 3N.

{\bf No. 15 = A952+69}\, The photograph in KKB85 shows a very low-surface
brightness object with clumpiness indicative of star formation.
It was resolved into stars with the 6-m. by Efremov et al. (1986).
Again the non-detection of this Im galaxy with the \nan\ radio
telescope does not necessarily imply this galaxy to be gas-poor. 
VLA synthesis observations in D-array by Yun et al.~(1994) in an
area of 1.5 $\Box\degr$ around the galaxy
M81 reveals a concentration of \HI at the position of this 
irrregular galaxy (their concentration II), suggestive
of an \HI mass of \MHI$=3.0\cdot 10^8$ \Msun\ (at the adopted
distance of 4 Mpc to the M81 group). The mean velocity of this
gas clump is about 100 \kms\ with a width of about 30 \kms\
(see also van der Hulst 1977, and Appleton et al. 1981).
With the single dish observation obtained here this signal 
cannot be resolved from the much stronger emission of M81.

{\bf No. 16 = Kar 61 = A0961+68 = Mailyan 47}\,
A photograph is given in Bertola \& Maffei (1974), an isodensity map
in KKB85, the luminosity profile and structural parameters in KKBRF87.
This is a typical example of a low-surface brightness dwarf spheroidal
system and the non-detection at \nan\ is not surprising. However,
the analysis by Appleton et al. suggest that this dwarf spheroidal is 
associated with a clump of \HI gas of the order of \MHI$ = 2.6\cdot
10^7$ \Msun\ with a mean velocity of about $V_{\rm hel} = -87$ \kms and a 
linewidth of $\sim$35 \kms. This clump is visible in the more detailed
\HI map by Yun et al.~(1994), but it looks more like the end of a
spiral arm than a distinct \HI\ concentration. In either case, the
expected signal would not be sufficiently strong to stand out over
the M81--M82--NGC3077\, \HI complex (cf. also Fig.~3).

{\bf No. 19 = Garland}\, 
Deep photographs, spectra and a detailed discussion of this dwarf
irregular are presented in Karachentseva et al. (1985b).
The total extent of the object is about 6$' \times 4'$ and its 
centre lies about \am{4}{5} from that of NGC 3077, which has an about 
55 \kms\ lower redshift. It shows various knots of star formation and
lies in the extension of the \HI bridge connecting NGC 3077 to M81.
Garland may be at an intermediate stage in the conversion of a tidal 
tail or bar into an ordinary dwarf galaxy as a result of the
interaction between M81 and NGC 3077. Its internal motions are of 
order 55 \kms\ and its dynamical age is estimated to be about 10$^8$ 
years only. The \HI\ distribution within NGC 3077 shows an extension
to the SE. This might be associated with Garland. The signal in our 
spectrum primarily originates from NGC 3077 ($V$ = 14 \kms, 
$\Delta V$ = 93 \kms). 

{\bf No. 20 = Kar 5N}\, The faintest object in our sample ($m_{\rm B}$=
18.1). It is round and without structural details (cf. KKB85). It is
outside the \HI complex around M81, hence any gas emission could only
reside within the narrow velocity range of the Galactic gas.

{\bf No. 21 = DDO 71 = Kar 63 = UGC 5428 = Mailyan 49}\, 
The photograph in KKB85 finds DDO 71 to be a circular,
structureless, very low-surface brightness object. It
was not detected in the earlier \HI work by
FT81 with rms noise of 18 mJy, or at 
Green Bank (STMM92) with an rms of 8.5 mJy.
It is not found in co-added IRAS data either 
(Melisse \& Israel 1994). Based on its optical radial velocity
($V = -126$ \kms) and the Galactic gas emission as seen
in the spectrum, its \HI emission -- if present -- should have 
been measurable.

{\bf No. 23 = Kar 64 = UGC 5442 = Mailyan 50}\, 
A photograph and isodensity map are given in KKB85. 
Not detected previously by STMM92.

{\bf No. 24 = DDO 78}\,
The photograph in KKB85 shows a circular, very low-surface brightness 
object. Not detected by FT81.

{\bf No. 25 = U 5658 = Mailyan 53}\, 
The overexposed photograph in KKB85 shows
a high-surface brightness object with a regular elliptical 
outline, partially resolved into knots surrounded by diffuse material.
This low flux object was not detected in the earlier 
less sensitive \HI\ surveys (FT81 \& STMM92). 
The NED search finds two galaxies in the 1.5 times
the HPBW area, but they do not appear to be related to our 
\HI line detection: CGCG 333-030, a 15.7 mag object of unknown
morphological type and redshift,
at 8\farcm9 distance ($\alpha = 10^{\rm h}24^{\rm m}24^{\rm s}$, 
$\delta = 71\degr21\arcmin00\arcsec$), and UGC 5645, a 16$^{\rm th}$ mag 
SBb spiral at a redshift of 10,490 \kms, at 11\farcm2 distance
($\alpha = 10^{\rm h}23^{\rm m}16\fs1$, $\delta =71\degr40\arcmin25\arcsec$).

{\bf No. 27 = DDO 82 = UGC 5692}\,
The photograph in KKB85 shows a dwarf spiral galaxy with a very 
distorted spiral structure. Its optical spectrum shows bright \Ha\ 
and $[$S{\sc ii}$]$ lines (Karachentsev \& Karachentseva 1984)
yielding a recesson velocity of $V$ = 40 \kms.
Note, that the radial velocity of 180 \kms listed in the NED database
and elsewhere is wrong, since the solar motion has been corrected for 
twice (Karachentsev et al. 1994). A distance of 4.5: Mpc was derived from 
photometry of its brightest stars (Karachentsev et al. 1994).
Its morphology, the distance and its detection at 60 and 100 $\mu$m 
in co-added IRAS data (Melisse \& Israel 1994) strongly suggest
this galaxy to be \HIshort-rich. But it was not detected in the
present \nan\ survey, nor in the earlier surveys 
(FT81, STMM92), nor in a short
Westerbork interferometer observation (Kamphuis et al. 1996).
This can only be explained, if the gas is at the same velocity as 
the optically obtained velocity ($V$ = 40 \kms) rendering
detection impossible due to Galactic \HI emission.

{\bf No. 28 = Kar 6N}\,
The photograph and isodensity map in KKB85
show it to be a low-surface brightness object with an axial ratio 
of 0.5, hardly any luminosity gradient and no structural detail.

{\bf No. 30 = UGCA 220 = Mailyan 58}\,
The KKB85 photograph reveals a diffuse, lumpy object
with patches of different surface brightness suggestive of
star formation and presence of \HI gas. This galaxy was not detected
at \nan, nor earlier by FT81. It may be part of the clustering
of peculiar low-surface brightness objects, 5 of which were included in 
our present observing program. 

{\bf No. 33 = Kar 7N}\,
The KKB85 photograph shows a diffuse, patchy faint object. Belongs to the
group of unusual objects discovered by B\"orngen et al.~(1984).

{\bf No. 36 = Kar 73}\, 
Photographs in KKB85 and Tikhonov \& Karachentsev
(1993) find this galaxy to be resolved into separate knots (stars 
and \HII\ regions?) superimposed on a diffuse, very low-surface 
brightness background. Tikhonov \& Karachentsev determine a 
photometric distance of 4.0 Mpc from the brightest stars. 
VLA \HI\ imaging (Westpfahl \& Puche 1994) shows a structure resembling
an edge-on disc with a central hole; the velocity field is too complex
to discern systematic rotation.

{\bf No. 37 = Anon 5}\, Belongs to the group of peculiar objects.
Photograph in B\"orngen et al. (1984).

{\bf No. 38 = Anon 6}\, Belongs to the group of peculiar objects.
Photograph in B\"orngen et al. (1984). 

{\bf No. 39 = Kar 8N}\, Diffuse, patchy object of large angular size, at
the detection limit of the 6m photograph in KKB85. Belongs to the group of
unusual objects reported by B\"orngen et al.~(1984).

{\bf No. 40 = Anon 7}\, Belongs to the group of peculiar objects.
Photograph in B\"orngen et al. (1984). 

{\bf No. 41 = Kar 74}\,
The photograph (cf. KKB85) shows a featureless object of regular 
elliptical shape.

\section{Discussion}
\subsection{\HIit\ line detections}
We have observed 23 dwarf galaxy candidate members of the M81 group of
galaxies, including 5 peculiar objects of unknown nature. Of 
the 18 with given morphology, half are early types (dE or dS0: and one
dE/Im:), and half are late types (Im or Sm). The latter 9 objects
should in principle be gas-rich, 
hence detectable at \nan\ with the sensitivity obtained here.

However, in the present survey only 3 objects were detected in the 
\HI line. The strong signal in the spectrum of No. 11 (Kar 59) at 
1323 \kms\ originates from a nearby bright spiral. 
None of the 3 detected dwarfs seem to be a member of the M81 group
of galaxies, however, as their redshifts of 632, 570 and 1159 \kms\ 
clearly show them to be behind the group. If we calculate distances
directly from their recession velocites corrected to the LSR following
the precepts given in Sandage \& Tammann (1987), and a Hubble constant 
of $H_0$ = 50 \kmsMpc, the galaxies UGC 4998, Kar 1N and UGC 5658 are
at distances of 15.9, 14.7 and 26.6 Mpc and have \HI masses of 0.05, 
2.0 and 2.5$\cdot 10^{8}$\Msun. The resulting \HI mass-to-blue light 
ratios are 0.05, 0.91 and 0.22 \MsunLBsun, respectively. 

The profile of Kar 1N is narrow ($\Delta V_{50}$ = 61 \kms) and clearly
Gaussian shaped, that of UGC 4998 is a bit broader (89 \kms) and more 
flat-topped, while that of UGC 5658 ($\Delta V_{50}$ = 127 \kms) least 
resembles a typical dwarf line profile.

The two detected objects classified as magellanic irregulars (Kar 1N 
and UGC 5658) have quite discrepant \MHILB\ ratios 
(0.91 and 0.22 \MsunLBsun) for their similar morphological type. In particular,
the \MHILB\ ratio of 0.22 for UGC 5658 seems very low for an
irregular dwarf, whereas the extremely low \MHILB\ ratio of
0.05 \MsunLBsun\ for the dS0: galaxy UGC 4998 seems quite consistent with 
the upper limit of about 0.15 found for morphologically pure lenticulars
(Knapp et al.~1989).

For comparison, the global optical and \HI line properties 
of all 11 previously detected M81-group dwarf galaxies are given Table 3. 
The values are taken from the Huchtmeier \& Richter (1989) and Schmidt \&
Boller (1992) compilations, with the exception of the integrated
fluxes for the systems M81, M82, and NGC 3077 interconnected in \HIshort, 
as well as Ho\,{\sc ix}, which are adopted from the VLA synthesis analysis by 
Yun et al.~(1994).

%
%
\begin{table*}
\bigskip
\small
\begin{tabular}{rlcclrrrrrc}
\multicolumn{11}{l}{{\bf Table 3:} Members of the M81 group detected
 previously in \HI} \\
\smallskip \\
\hline
\vspace{-3 mm} \\
& Ident. & R.A. & Dec. & Type & Diam & $B_{\rm T}$ & $V_{\rm hel}$ 
& $\int$$S$d$V$ &
  $\Delta V_{20}$ & \MHILB\ \\
 & & (h\,\, m\,\, s) & (\degr\,\, \arcmin\,\, \arcsec) 
& & $(\,'\,)$ & {\footnotesize (mag)} & 
  {\footnotesize (km/s)} & {\footnotesize (Jy km/s)} 
  &  {\footnotesize (km/s)} & {\footnotesize (\MsunLBsun)} \\
\vspace{-3 mm} \\
\hline
\vspace{-2 mm} \\
 1. & Ho\,{\sc ii}    & 08 13 53.5 & 70 52 13 & Im     &  7.9 & 10.2 & 158 &
 359.7 &  79  & 0.59 \\  
 2. & Kar 52   & 08 18 43.0 & 71 11 25 & Im     &  1.3 & 14.2 & 114 &
 3.7 &  38  & 0.25 \\ 
 3. & DDO 53   & 08 29 33.3 & 66 21 08 & Im     &  1.5 & 13.6 &  19 &  23.7 
&  46: & 0.89 \\  
 4. & UGC 4483 & 08 32 07.0 & 69 57 16 & Im     &  1.1 & 13.9 & 157 &
  3.1 &  70 & 0.16 \\  
 7. & Ho\,{\sc i}     & 09 36 00.9 & 71 24 55 & IABm   &  3.6 & 12.2 & 136 &  49.0 
&  45 & 0.50 \\
 9. & NGC 2976 & 09 43 11.5 & 68 08 45 & SAcp   &  5.9 & 10.9 &   3 &  63.6 
& 159 & 0.20 \\
13. & M81      & 09 51 27.3 & 69 18 08 & Sb     & 26.9 &  7.9 & --34 &
 859.9 & 464 & 0.19 \\
14. & M82      & 09 51 43.6 & 69 55 00 & Amorph & 11.2 &  9.3 & 203 & 245.2 
& 290 & 0.20 \\
17. & Ho\,{\sc ix}    & 09 53 28.0 & 69 16 53 & Im     &  2.5 & 13.5 &  46 &
 94.8 & 120: & 3.63 \\
18. & NGC 3077 & 09 59 21.9 & 68 58 33 & Am     &  5.4 & 10.6 &  14 & 212.5 
&  93 & 0.57 \\
22. & UGC 5423 & 10 01 25.3 & 70 36 27 & BCD    &  0.9 & 13.8 & 343 &   2.3 
&  80 & 0.11 \\
26. & IC 2574  & 10 24 41.3 & 68 40 18 & Sm     & 13.2 & 11.0 &  47 & 442.5 
& 126 & 1.52 \\
31. & DDO 87   & 10 46 17.0 & 65 47 40 & Im:    &  2.4 & 14.9 & 338 &  18.9 
&  80 & 2.35 \\
\vspace{-2 mm} \\
\hline \\
\end{tabular}
\normalsize
\end{table*}

\subsection{The non-detections and confusion problems}
The average 3$\sigma$ upper limit of 13 mJy and an assumed line 
width of 50 \kms imply an upper limit of 0.65 Jy \kms\ to the 
\HI line flux for the objects not detected in our survey. This implies 
an upper mass limit of $ 2 \cdot 10^{6}$ \Msun\ (and even less for
the ``anonymous'' low-surface brightness objects of unknown nature)
at the adopted distance of 4 Mpc to the M81 group, and upper limits
to the \MHILB\ ratios of 0.1 and 0.6 \MsunLBsun\ for a 15$^{\rm th}$,
respectively a 17$^{\rm th}$ magnitude galaxy.

However, the non-detections are unlikely due to a lack of gas in these 
dwarfs. Inherent to \HI line searches for nearby objects is 
always the problem of confusion by or with strong Galactic \HI lines,
i.e. emission lines of external galaxies being lost among strong
Galactic lines, or being interpreted as part of the Galactic 
emission (see, e.g., the case of the discovery of Dwingeloo 1 by 
Kraan-Korteweg et al. 1994).
We have noted the velocity ranges dominated by Galactic confusion for 
each object in Table 2, i.e., the range in which we estimate that 
the profile of a typical dwarf with a peak intensity of 10 mJy would 
not be recognized as such. The average range is about --150 to 115 \kms\
with about $\pm$50 \kms\ for the 2 narrowest profiles. As the mean velocity
of the 18 members of the M81 group with known redshifts listed in Table 1
is $<V>$ = 101 \kms, with a dispersion of $\sigma$ = 114 \kms, this means 
that,
statistically speaking, more than half of them would be lost among strong 
Galactic \HI lines if they were all gas-rich. In fact, 5 of the galaxies
observed at \nan\ have known optical redshifts (see Table 2), 
which {\em all}\/
fall within the velocity range obscured by Galactic \HIshort.

In the central part of this group, i.e. in an area of about 1.5 
$\Box\degr$ around the galaxy M81, the confusion problem is even 
worse. Here, M81, M82, NGC 3077, and -- according to Appleton et
al.~(1981, their Fig.~2) -- also NGC 2976 are embedded in a common, 
very extended \HI cloud with \HI bridges connecting the major 
galaxies, and distinct \HI clumps, some of which coincide with 
optically identified dwarfs. 

Seven dwarfs (Kar 2N, Kar 3N, A9562+68, Kar 61, Ho {\sc ix}, Garland, and 
Kar 5N) 
reside in this area where the strong \HI emission (within about --280
to +355 \kms) from the larger interacting galaxies makes the
detection of gas-rich dwarfs extremely difficult (local estimates for the
confusion range are given for each dwarf candidate in the last column 
of Table 2). So far, only the dwarf Ho {\sc ix} has been 
unambigously associated with \HIshort. None of the 6 dwarfs observed at 
\nan\ were detected. If we regard Fig.~1 of Yun et al.~(1994), however, 
it seems clear that the irregular galaxy A952+69 must be associated 
with the second \HI concentration visible in that image (the first being
Ho {\sc ix}). Appleton et al.~(1981) even suggest that the \HI concentration 
in the southern tip of the spiral arm of M81 visible in \HI is due to Kar 61 
(No.16) and that the south eastern extension in the \HI distribution 
of NGC 3077 might be due to the starforming galaxy Garland.

Even with the above indications for gas in three of
the central dwarf galaxies, single dish observations cannot resolve 
these signals from the stronger emission of their dominant companions 
-- or the Galaxy. In case of the two objects with the broadest 
confusion range in the \nan\ beam (No. 15 = A952+69, --270 to +355 \kms,
and No. 16 = Kar 61, --260 to +190 \kms), we have obtained
shorter observations pointed to the galaxy's centre and one HPBW (4$'$) 
due East and one due West, in order to verify whether part of the
signal could be assigned to the dwarf candidate in question.
A comparison of these on and off-source profiles (Fig.~3) confirms 
the signal to originate from our Galaxy and the \HI complex around
M81, as they have about the same velocity extent at and near each
galaxy, whose optical diameter is considerably smaller than the 
beam width. 

%
 \begin{figure*} 
\hfil \epsfxsize 18cm \epsfbox{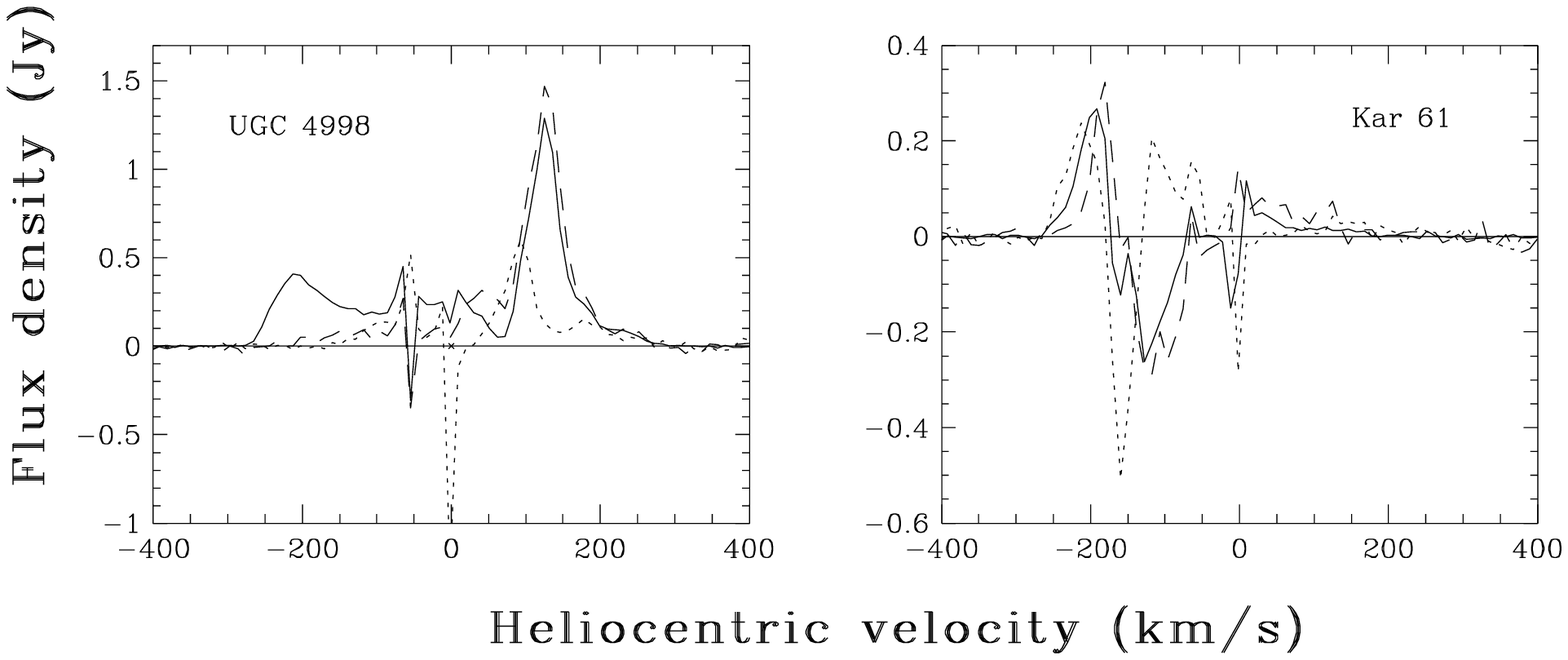}\hfil
 \caption[]{A comparison of spectra obtained exactly at the optical
   centre of the two objects (No. 15, A952+69 and No. 16, Kar 61) with
offsets of one beam width (4$'$) due East and West.
The fact that the line profiles are so similar makes it impossible to
disentangle the expected much fainter contribution of the dwarf from
the signal which is due to the Galaxy and the M81 complex.
 }
 \end{figure*} 

Not even the detailed \HI maps from synthesis observations can
resolve these problems. The assignment of individual \HI concentrations
(or filaments) to optically identified dwarf galaxies 
remains ambiguous within this highly active region where the gas might
have been swept away from the currently visible starforming dwarfs.

Here spectroscopy, in particular the determination of the recession
velocities from emission lines in starforming galaxies, might be the
only approach to learn more about the kinematics in the M81 group -- not only
in the central part of the group, but also within the
larger group boundary.

\subsection{Conclusion}

Three new detections of dwarf candidates in
the M81 group have disproven these objects to be members of the
M81 group. Due to confusion problems the non-detections 
do not a priori imply that the other observed dwarf candidates 
are gas-poor, although the high-sensitivity observations clearly 
demonstrate that they cannot be gas-rich dwarfs just beyond the M81
group. In order to study the dynamics of the M81 group in further
detail, in particular the behaviour of the dwarfs within the
gravitational potential well of the group as such and the massive
galaxies within it, spectroscopic observations of the remaining
dwarf candidates without velocity information are required.

\acknowledgements{ 
We would like to thank Drs. L. Bottinelli, L. Gouguenheim and G. Theureau
for making their flux density calibration data available to us, and V. Etieve
for her help with the illustrations.
The \nan\ Radio Observatory is the Unit\'e Scientifique \nan\
of the Observatoire de Paris, associated as Unit\'e de Service et de
Recherche (USR) No. B704 to the French Centre National de Recherche 
Scientifique (CNRS). The Observatory also gratefully acknowledges the 
financial support of the R\'egion Centre in 
France. The research by RCKK is being supported with an EC grant.
This research has made use of the NASA/IPAC Extragalactic Database (NED)   
which is operated by the Jet Propulsion Laboratory, California Institute   
of Technology, under contract with the National Aeronautics and Space      
Administration.                                                            
 }

\end{document}